\documentclass[runningheads]{llncs} %

\usepackage[utf8]{inputenc}
\usepackage[T1]{fontenc}
\usepackage[english]{babel}
\usepackage{microtype}      %
\usepackage{csquotes}

\usepackage{amsmath}
\usepackage{amssymb}
\usepackage{bm}
\usepackage{textcomp}
\usepackage{booktabs}
\usepackage{pifont}  %

\usepackage[pdftex]{graphicx}
\usepackage[table]{xcolor}
\usepackage{hyperref}

\newcommand{\ourmod}{DAFT}
\newcommand{\balpha}{\ensuremath\bm{\alpha}}
\newcommand{\bbeta}{\ensuremath\bm{\beta}}

\newcommand{\cmark}{\ding{51}}%
\newcommand{\xmark}{\ding{55}}%
\newcommand{\ubold}[1]{\fontseries{b}\selectfont#1}

\begin{document} %
\title{Combining 3D Image and Tabular Data via the Dynamic Affine Feature Map Transform\thanks{S.~P{\"{o}}lsterl and T.N.~Wolf -- These authors contributed equally to this work.}}
\titlerunning{The Dynamic Affine Feature Map Transform}
\author{Sebastian P{\"{o}}lsterl \and
  Tom Nuno Wolf \and
  Christian Wachinger} %
\authorrunning{S.~P{\"{o}}lsterl, T.~N.~Wolf and C.~Wachinger} %
\institute{Artificial Intelligence in Medical Imaging (AI-Med),\\
    Department of Child and Adolescent Psychiatry,\\
    Ludwig-Maximilians-Universit{\"{a}}t, Munich, Germany%
} %
\maketitle              %
\begin{abstract}
  Prior work on diagnosing Alzheimer's disease from magnetic resonance images of the brain
  established that convolutional neural networks (CNNs) can leverage
  the high-dimensional image information for classifying patients.
  However, little research focused on how these models can utilize
  the usually low-dimensional tabular information, such
  as patient demographics or laboratory measurements.
  We introduce the Dynamic Affine Feature Map Transform (DAFT),
  a general-purpose module for CNNs that dynamically
  rescales and shifts the feature maps of a convolutional layer,
  conditional on a patient's tabular clinical information.
  We show that DAFT is highly effective in combining 3D image and
  tabular information for diagnosis and
  time-to-dementia prediction, where it outperforms competing
  CNNs with a mean balanced accuracy
  of 0.622 and mean c-index of 0.748, respectively.
  Our extensive ablation study provides valuable insights into
  the architectural properties of DAFT.
  Our implementation is available at \url{https://github.com/ai-med/DAFT}.
\end{abstract} %
\section{Introduction}

In recent years, deep convolutional neural networks (CNNs) have become
the standard for classification of Alzheimer's disease (AD) from
magnetic resonance images (MRI) of the brain
(see e.g. \cite{Ebrahimighahnavieh2020,Wen2020} for an overview).
CNNs excel at extracting high-level information about the neuroanatomy
from MRI.
However, brain MRI only offers a partial view on the underlying changes causing
cognitive decline.
Therefore, clinicians and researchers often rely on tabular
data such as patient demographics, family history, or laboratory
measurements from cerebrospinal fluid for diagnosis.
In contrast to image information, tabular data is typically
low-dimensional and individual variables capture rich clinical knowledge.

Due to image and tabular data being complementary to each other, it is
desirable to amalgamate both sources of information in a single neural network
such that one source of information can inform the other.
The effective integration is challenging, because the dimensionality mismatch
between image and tabular data necessitates an architecture where the capacity
required to summarize the image is several orders of magnitude higher than the
one required to summarize the tabular data.
This imbalance in turn implicitly encourages the network to focus on
image-related parameters during training, which ultimately can result in
a model that is only marginally better than a CNN using the
image data alone~\cite{Pelka2020}.
Most existing deep learning approaches integrate image and tabular data
na{\"i}vely by concatenating the latent image representation with the tabular data
in the final layers of the network~\cite{Esmaeilzadeh2018,Hao2019,Kopper2020,Liu2019,Mobadersany2018,Poelsterl2019}.
In such networks, the image and tabular parts
have only minimal interaction and are limited in the way one part can inform the other.
To enable the network to truly view image information in the context of the
tabular information, and vice versa, it is necessary to increase the network's capacity
and interweave both sources of information.

We propose to increase a CNN's capacity
to fuse information from a patient's 3D brain MRI
and tabular data via the Dynamic Affine Feature Map Transform (\ourmod).
\ourmod\ is a generic module that can be integrated into any
CNN architecture that establishes a two-way exchange of information between
high-level concepts learned from the 3D image and the tabular biomarkers.
\ourmod\ uses an auxiliary neural network to dynamically incite or repress
each feature map of a convolutional layer conditional on both
image \emph{and} tabular information.
In our experiments on AD diagnosis and time-to-dementia prediction,
we show that \ourmod\ leads to superior predictive performance than
using image or tabular data alone, and that it outperforms previous approaches that
combine image and tabular data in a single neural network by a large margin.

\section{Related Work}

A na{\"i}ve approach to combine image and tabular data
is to first train a CNN on the image data, and
use its prediction (or latent representation) together with
tabular data in a second, usually linear, model.
This way, the authors of \cite{Li2019} combined
regions of interest extracted from brain MRI with
routine clinical markers to predict progression to AD.
Since image descriptors are learned independently of the clinical
markers, descriptors can capture redundant information,
such as a patient's age, instead of complementing it.
This is alleviated when using a single network that
concatenates the clinical information with the latent image representation
prior to the last fully connected (FC) layer,
which has been done in \cite{Hao2019} with
histopathology images, genomic data, and demographics for survival prediction,
and in \cite{Kopper2020,Poelsterl2019} with hippocampus shape and clinical
markers for time-to-dementia prediction.
The disadvantage of this approach is that tabular data
only contributes to the final prediction linearly.
If concatenation is followed by a multilayer perceptron (MLP), rather
than a single FC layer,
non-linear relationships between image and tabular data can be captured.
This was applied by the authors of~\cite{Mobadersany2018} to learn from
digital pathology images and genomic data, and of
\cite{Esmaeilzadeh2018,Liu2019} to learn from brain MRI and
clinical markers for AD diagnosis.
Closely related to the above, the authors of \cite{ElSappagh2020,Li2020,Spasov2019}
use an MLP on the tabular data
before concatenation, and on the combined representation after concatenation.
However, both approaches are restricted to interactions between the global
image descriptor and tabular data and do not support fine-grained
interactions.

In contrast,
Duanmu~et~al.~\cite{Duanmu2020} fused information in a multiplicative manner
for predicting response to chemotherapy.
They use an auxiliary network that takes
the tabular data and outputs a scalar weight for each feature map
of every other convolutional layer of their CNN.
Thus, a patient's tabular data can amplify or repress the contribution
of image-derived latent representations at multiple levels.
The downside of their approach is that the number of weights in
the auxiliary network scales quadratically with the depth of the CNN,
which quickly becomes impracticable.
The Feature-wise Linear Modulation (FiLM) layer,
used in visual question-answering, is the most similar to our approach~\cite{Perez2018}.
FiLM has an auxiliary network that takes the text of the question and
outputs an affine transformation to scale and shift each feature map
of a convolutional layer.
In the medical domain, the only approach based on FiLM is for image segmentation
to account for lesion size or cardiac cycle phase~\cite{Jacenkow2020}.
In contrast to the above, we focus on disease prediction and utilize tabular information
that is complementary to the image information,
rather than describing image contents or semantics.
Moreover, our proposed
\ourmod\ scales and shifts feature maps of a convolutional layer
conditional on both image \emph{and} tabular data.

\section{Methods}

\begin{figure}[t]
  \centering
  \includegraphics[width=\textwidth]{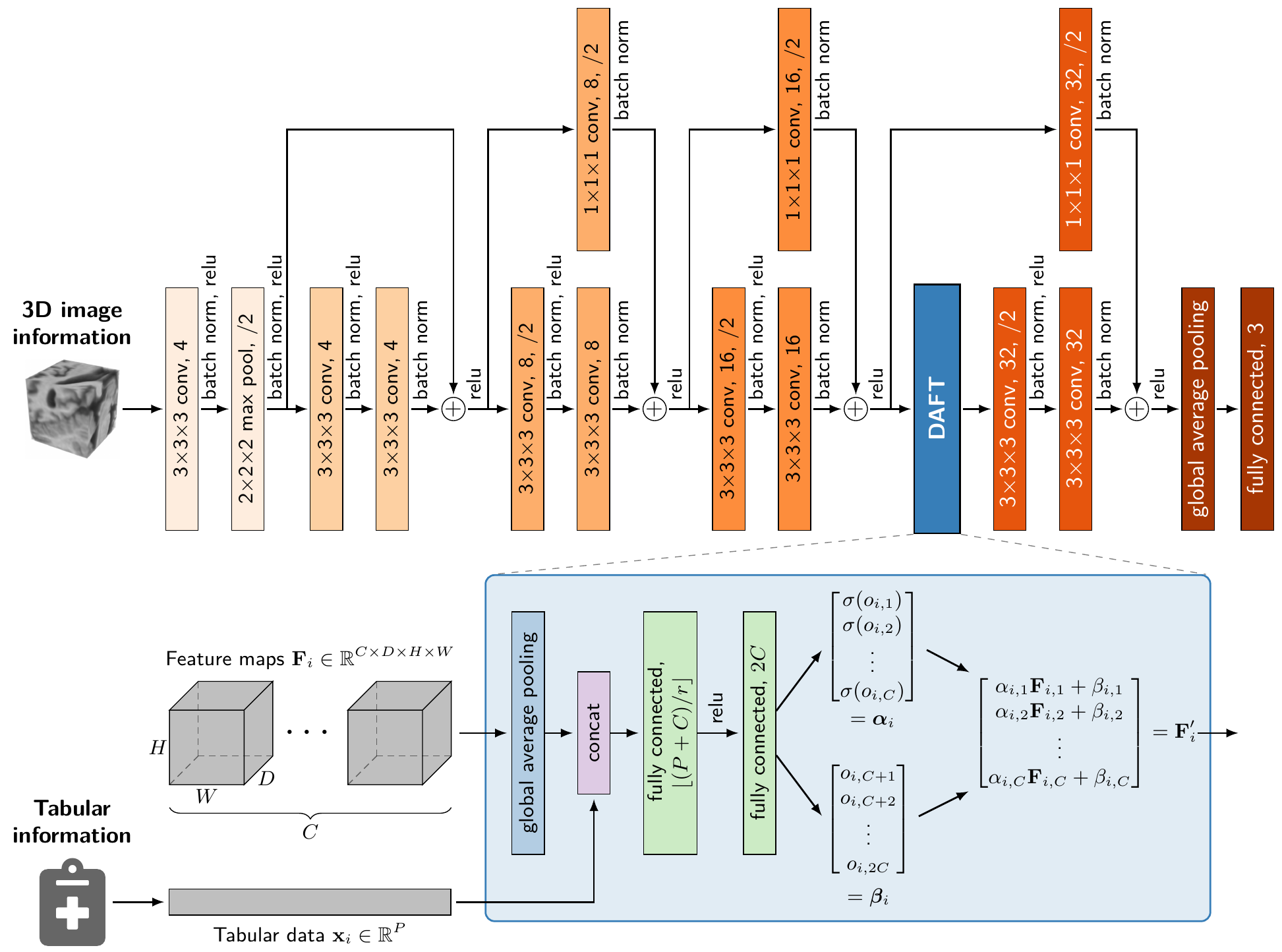}
  \caption{\label{fig:network}%
  Our proposed network architecture with the Dynamic Affine Feature Map Transform (\ourmod)
  in the last residual block.
  \ourmod\ combines
  $\mathbf{F}_i$, a $C \times D \times H \times W$ tensor of $C$
  feature maps from a convolutional layer, and $\mathbf{x}_i \in \mathbb{R}^P$, a vector of tabular data,
  to affinely transform the $C$ feature maps
  via scales $\bm{\alpha_{i}}$ and shifts $\bm{\beta}_{i}$,
  where $r=7$.}
\end{figure}

We are seeking a CNN that utilizes high-dimensional 3D image information
and seamlessly accounts for complementary low-dimensional tabular information
in its predictions.
We use a ResNet~\cite{He2016} architecture and
achieve tight integration of both sources of information
by dynamically scaling and shifting the feature maps of a 3D convolutional layer,
conditional on a patient's image and clinical tabular information.
Since tabular information often comprises demographics and summary measures
that describe the patient's state as a whole,
we require a level exchange of information between
tabular and image data.
Therefore, we propose to affinely transform the output
of a convolutional layer in the last residual block,
which is able to describe the image in terms of high-level concepts
rather than primitive concepts, such as edges.
Fig.~\ref{fig:network} summarizes our network.

For the $i$-th instance in the dataset, we
let $\mathbf{x}_i \in \mathbb{R}^P$ denote the tabular clinical information, and
$\mathbf{F}_{i,c} \in \mathbb{R}^{D \times H \times W}$ denote the $c$-th output (feature map)
of a convolutional layer based on the $i$-th volumetric image ($c \in \{1,\ldots,C\}$).
We propose to incite or repress high-level concepts learned from
the image, conditional on the image and tabular data.
To this end, we learn the Dynamic Affine Feature Map Transform (\ourmod),
with scale $\alpha_{i,c}$ and offset $\beta_{i,c}$:
\begin{equation}
  \mathbf{F}^\prime_{i,c} = \alpha_{i,c} \mathbf{F}_{i,c} + \beta_{i,c},
  \qquad %
  \alpha_{i,c} =  f_c(\mathbf{F}_{i,c}, \mathbf{x}_i),\qquad
  \beta_{i,c} =  g_c(\mathbf{F}_{i,c}, \mathbf{x}_i),
\end{equation}
where $f_c$, $g_c$ are arbitrary functions that map the
image and tabular data to a scalar.
We model $f_c$, $g_c$ by a single auxiliary neural network $h_c$
that outputs one $\balpha$, $\bbeta$ pair, which we refer to as
\ourmod\ (see Fig.~\ref{fig:network}).
\ourmod\ first creates a
bottleneck by global average pooling of the image feature map,
concatenating the tabular data,
and squeezing the combined vector by a factor $r$
via an FC layer.
Next, both vectors are concatenated and fed to an MLP that
outputs vectors $\balpha_i$ and $\bbeta_i$;
following~\cite{Hu2020}, all FC layers do not have bias terms.
Note that we only apply an activation function $\sigma(\cdot)$ to the scale $\balpha$,
but not the offset $\bbeta$ -- we explore linear, sigmoid, and tanh
activations in our experiments.
The proposed \ourmod\ is computationally efficient, because it
does not depend on the number of instances in the dataset, nor
the spatial resolution of the feature maps. \ourmod\ can dynamically
produce a scale factor $\alpha_{i,c}$ and offset $\beta_{i,c}$
conditional on a patient's specific image and tabular information
due to parameter sharing.
Finally, note that our principal idea of \ourmod\ is not restricted
to the CNN in Fig.~\ref{fig:network}, but can be integrated
in any type of CNN.

\section{Experiments}

\begin{table}[tb]
  \centering
  \caption{\label{tab:datasets}%
  Dataset statistics.}
  \begin{scriptsize}
  \begin{tabular}{lrccl}
    \toprule
    \multicolumn{1}{c}{Task} & Subjects & Age & Male & Diagnosis \\
    \midrule
    Diagnosis   & 1341 & $73.9 \pm 7.2$ & 51.8\% & Dementia (19.6\%), MCI (40.1\%), CN (40.3\%) \\
    Progression & 755 & $73.5 \pm 7.3$ & 60.4\% & Progressor (37.4\%), median follow-up time 2.01 years \\
    \bottomrule
  \end{tabular}
  \end{scriptsize}
\end{table}

We evaluate \ourmod\ on two tasks using T1 brain MRI from
the Alzheimer's Disease Neuroimaging Initiative~\cite{Jack2008}:
(i) diagnosing patients as cognitively normal (CN),
mild cognitive impaired (MCI), or demented,
and (ii) predicting the time of dementia onset for MCI patients.
We formulate the diagnosis task as a classification problem,
and the time-to-dementia task as a survival analysis problem,
i.e., dementia onset has only
been observed for a subset of patients, whereas the remaining patients
remained stable during the observation period (right censored time of onset).
Table~\ref{tab:datasets} summarizes the datasets.

\subsection{Data Processing}
We first segment scans with
FreeSurfer~\cite{Fischl2012} and extract a region of interest of size $64^3$
around the left hippocampus, as it is known to be strongly affected by AD~\cite{Frisoni2008}.
Next, we normalize images following the minimal pre-processing pipeline
in~\cite{Wen2020}.
Tabular data comprises 9 variables: age, gender, education, ApoE4,
cerebrospinal fluid biomarkers A$\beta_{42}$, P-tau181 and T-tau,
and two summary measures derived from
18F-fluorodeoxyglucose and florbetapir PET scans.
To account for missing values, we adopt an approach similar to \cite{Jarrett2020}
by appending binary variables indicating missingness
for all features, except age, gender, and education which are always present.
This allows the network to use incomplete data and learn from missingness patterns.
In total, tabular data comprises $P=15$ features.
To avoid data leakage due to confounding effects of age and sex~\cite{Wen2020},
data is split into 5 non-overlapping folds using only baseline visits
such that diagnosis, age and sex are balanced across folds~\cite{Ho2007}.
We use one fold as test set and combine the remaining folds such that
80\% of it comprise the training set and 20\% the validation set.
For the diagnosis task, we extend the training set, but not validation or test, by including each patient's
longitudinal data ($3.49 \pm 2.56$ visits per patient).
For the time-to-dementia task, we only include patients with MCI at baseline
and exclude patients with bi-directional change in diagnosis such
that all patients remain MCI or progress to dementia.

\subsection{Evaluation Scheme}

We consider two unimodal baselines:
(i) a ResNet~\cite{He2016} that only uses the image information
based on the architecture in Fig.~\ref{fig:network}, but without \ourmod,
and (ii) a linear model using only the tabular information.%
\footnote{We explored gradient boosted models too, but did not observe any advantage.}
Moreover, we fit a separate linear model where the
latent image representation from the aforementioned ResNet
is combined with the tabular data (Linear model /w ResNet features).
The linear model is a logistic regression model for diagnosis, and
Cox's model for time-to-dementia prediction~\cite{Cox1972}.
As baselines, we evaluate three concatenation-based networks
with the same ResNet backbone as in Fig.~\ref{fig:network}.
In Concat-1FC, the tabular data is concatenated with
the latent image feature vector
and fed directly to the final classification layer. Thus, it models
tabular data linearly -- identical to the linear model baseline -- but
additionally learns an image descriptor, as in \cite{Hao2019,Kopper2020,Poelsterl2019}.
In Concat-2FC, the concatenated vector is fed to an FC bottleneck layer
prior to the classification layer, similar to~\cite{Esmaeilzadeh2018,Liu2019}.
Inspired by~\cite{Spasov2019}, 1FC-Concat-1FC feeds the tabular information
to an FC bottleneck layer before concatenating it with the latent image representation, as in Concat-1FC.
In addition, we evaluate the network introduced
by Duanmu et al.~\cite{Duanmu2020}, and the FiLM layer~\cite{Perez2018},
originally proposed for visual question answering, in place of \ourmod.
Our implementation of the \ourmod\ and competing methods is available at \url{https://github.com/ai-med/DAFT}.

\begin{table}[t]
  \centering
  \caption{\label{tab:prediction-performance}%
    Predictive performance for the diagnosis task (columns 4-5)
    and time-to-dementia task (columns 6-7).
    Values are mean and standard deviation across 5 folds.
    Higher values are better. I indicates the use of image data,
    T of tabular data, with L/NL denoting a linear/non-linear model.
    }\setlength{\tabcolsep}{.3em}
  \begin{scriptsize}
  \begin{tabular}{p{2.36cm}llccccc}%
    \toprule
    & & & \multicolumn{2}{c}{Balanced Accuracy} && \multicolumn{2}{c}{Concordance Index} \\
    & I & T & Validation & Testing && Validation & Testing \\
    \midrule
    Linear Model         & \xmark & L
    &       0.571 $\pm$ 0.024 &      0.552 $\pm$ 0.020 &
    &  0.726 $\pm$ 0.040 &  0.719 $\pm$ 0.077 \\
    ResNet               & \cmark & --
    &       0.568 $\pm$ 0.015 &      0.504 $\pm$ 0.016 &
    &  0.669 $\pm$ 0.032 &  0.599 $\pm$ 0.054 \\
    Linear Model\newline/w ResNet Features & {\color{gray}\cmark} & L
    &       0.585 $\pm$ 0.050 &      0.559 $\pm$ 0.053 &
    &  0.743 $\pm$ 0.026 &  0.693 $\pm$ 0.044 \\
    \hline
    Concat-1FC           & \cmark & L
    &       0.630 $\pm$ 0.043 &      0.587 $\pm$ 0.045 &
    &  0.755 $\pm$ 0.025 &  0.729 $\pm$ 0.086 \\
    Concat-2FC           & \cmark & NL
    &       0.633 $\pm$ 0.036 &      0.576 $\pm$ 0.036 &
    &  0.769 $\pm$ 0.026 &  0.725 $\pm$ 0.039 \\
    1FC-Concat-1FC       & \cmark & NL
    &       0.632 $\pm$ 0.020 &      0.591 $\pm$ 0.024 &
    &  0.759 $\pm$ 0.035 &  0.723 $\pm$ 0.056 \\
    Duanmu et al.~\cite{Duanmu2020} & \cmark & NL
    &       0.634 $\pm$ 0.015 &      0.578 $\pm$ 0.019 &
    &  0.733 $\pm$ 0.031 &  0.706 $\pm$ 0.086 \\
    FiLM~\cite{Perez2018}           & \cmark & NL
    &       0.652 $\pm$ 0.033 &      0.601 $\pm$ 0.036 &
    &  0.750 $\pm$ 0.025 &  0.712 $\pm$ 0.060 \\
    \ourmod                         & \cmark & NL
    &       0.642 $\pm$ 0.012 &      \ubold{0.622 $\pm$ 0.044} &
    &  0.753 $\pm$ 0.024 &  \ubold{0.748 $\pm$ 0.045} \\
    \bottomrule
    \end{tabular}
  \end{scriptsize}
\end{table}
\begin{table}[t]
  \centering
  \caption{\label{tab:runtime}%
  Training time for one epoch using a NVIDIA GeForce GTX 1080 Ti GPU.}%
  \begin{scriptsize}%
  \begin{tabular}{ccccccc}
    \toprule
    ResNet & Concat-1FC & Concat-2FC & 1FC-Concat-1FC & Duanmu~et~al.~\cite{Duanmu2020} &FiLM~\cite{Perez2018} & DAFT \\
    8.9\,s & 8.9\,s & 8.9\,s & 8.9\,s & 9.0\,s & 8.7\,s & 9.0\,s \\
    \bottomrule
  \end{tabular}
  \end{scriptsize}%
\end{table}

The networks Concat-2FC, 1FC-Concat-1FC, FiLM and \ourmod,
contain a bottleneck layer, which we set to 4 dimensions, which is
roughly one fourth of the number of tabular features.
For FiLM and \ourmod, we use the identity function $\sigma(x) = x$
in the auxiliary network for the scale $\alpha_{i,c}$.
In the diagnosis task, we minimize the cross-entropy loss. For
progression analysis, we account for right censored progression times
by minimizing the negative partial log likelihood of Cox's model~\cite{Faraggi1995}.
We use the AdamW optimizer for both tasks~\cite{Loshchilov2019}.
We train for 30 and 80 epochs in the diagnosis and progression task respectively,
and shrink the initial learning rate by 10 when 60\% has been completed and
by 20 when 90\% has been completed.
For each network, we optimize learning rate and weight decay on the
validation set using grid search with a total of $5 \times 3$ configurations.
\footnote{learning rate $\in \{ 0.03, 0.013, 0.0055, 0.0023, 10^{-3} \}$, weight decay $\in \{ 0, 10^{-4}, 10^{-2} \}$}
We report the performance on the test set
with respect to the best performing model on the validation set. %
For diagnosis, we use the balanced accuracy~(bACC; \cite{Brodersen2010}) to account
for class imbalance, and for time-to-dementia analysis, we use an
inverse probability of censoring weighted estimator of
the concordance index~(c-index; \cite{Uno2011}), which is identical to the area
under the receiver operating characteristics curve
if the outcome is binary and no censoring is present.

\section{Results}
\subsubsection{Predictive Performance.}

Table~\ref{tab:prediction-performance} summarizes the predictive performance
of all models.
For both tasks, we observe that the linear model using only the tabular data
outperforms the ResNet using only the image data.
This is expected as tabular data comprises amyloid-specific
measures derived from cerebrospinal fluid and PET imaging
that are known to become abnormal before changes in MRI are visible~\cite{Jack2013}.
Moreover, when learning image descriptors independently of the tabular clinical data
and combining both subsequently, the predictive performance does not
increase significantly (third row). This suggests that the learned image descriptor is not complementing
the clinical information.
In the diagnosis task, all Concat networks were successful
in extracting complementary image information, leading to an increase in
the average bACC by at least 0.024.
In the time-to-dementia task, the improvement in c-index is at most 0.01
over the linear model,
and when accounting for the variance, the improvement must be considered
insignificant.
The network by Duanmu et al.~\cite{Duanmu2020} generally performs worse than
the Concat networks and is outperformed by the linear model on
the time-to-dementia task too (0.013 higher mean c-index).
The FiLM-based network has a marginal lead over all Concat
networks on the diagnosis task (0.011 higher mean bACC), but
falls behind on the time-to-dementia task (0.013 lower mean c-index).
These results clearly demonstrate that Concat approaches cannot
achieve the level of integration to fully utilize the complementary nature
of image and tabular information, and that integrating tabular data with both
low and high-level descriptors of the image, as done by Duanmu et al.~\cite{Duanmu2020},
can severely deteriorate performance.
Our proposed \ourmod\ network is the only approach that excels at integrating
image and tabular data for both tasks by outperforming competing methods
by a large margin (0.021 higher bACC, 0.019 higher c-index).
Finally, table~\ref{tab:runtime} summarizes the training times of networks, which
shows that the runtime increase due to \ourmod\ is negligible.

\subsubsection{Ablation Study.}

To better understand under which settings \ourmod\ can best integrate
tabular data, we perform an ablation study with respect to
(i) its location within the last ResBlock, (ii) the activation function $\sigma$
for the scale $\alpha_{i,c}$, and (iii) whether one of scale and offset is sufficient.
Following~\cite{Perez2018}, we turn the parameters of batch normalization layers
that immediately precede the \ourmod\ off.
From Table~\ref{tab:ablation-study},
we observe that \ourmod\ is relatively robust to the choice of location.
Importantly, \ourmod\ outperforms all Concat networks on the diagnosis task,
irrespective of its location.
For the progression task, only the location before the first ReLU leads
to a performance loss.
Regarding the type of transformation, the results for diagnosing show
that scaling is more essential than shifting, but removing any of them
comes with a decisive performance drop of at least 0.013 bACC.
For progression analysis, the \ourmod's capacity seems
to be sufficient if one of them is present.
Finally, leaving the scale parameter unconstrained, as in the proposed
configuration, is clearly beneficial for diagnosis, but for progression
analysis constraining the scale leads to an increase in mean c-index.
Only two configurations of \ourmod\ are outperformed by Concat baselines,
which highlights its robustness.
Moreover, the optimal configuration differs between tasks,
hence, further gains are possible when optimizing
these choices for a given task.

\begin{figure}[t]
  \centering
  \includegraphics[width=\textwidth]{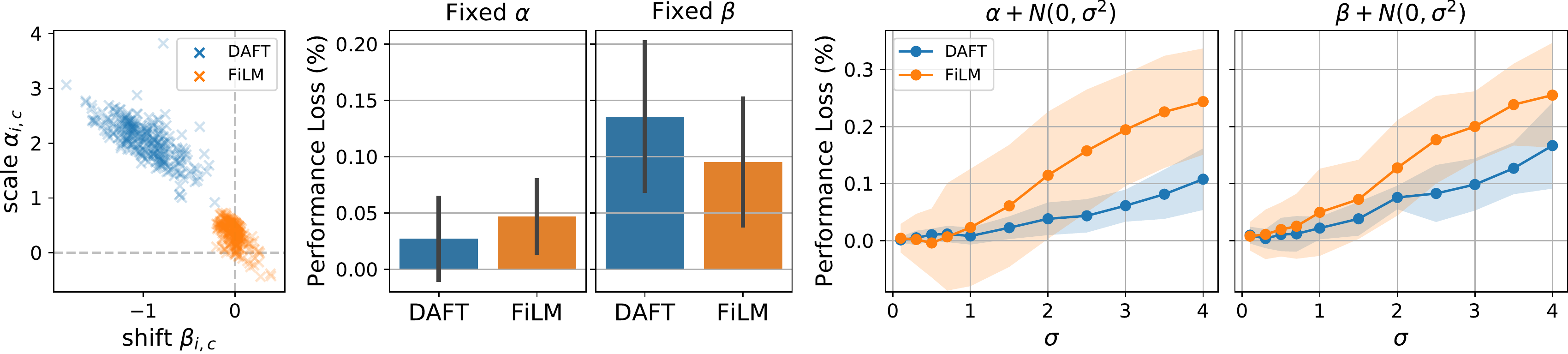}
  \caption{\label{fig:alphabeta}%
    Left: Scatter plot of $\alpha_{i,c}$ and $\beta_{i,c}$ for one feature map.
    Middle: Performance loss when setting $\balpha$ or $\bbeta$ to its mean.
    Right: Performance loss when distorting $\balpha$ or $\bbeta$.}
\end{figure}

\paragraph{Impact of $\balpha,\bbeta$.}
To compare the impact of $\balpha$ and $\bbeta$,
we run test time ablations on the fully trained models from the diagnosis task by modifying
$\balpha$ or $\bbeta$ during inference.
Fig.~\ref{fig:alphabeta}~(left) shows that the range of $\balpha$ and $\bbeta$
is higher for \ourmod\ than for FiLM, which uses $\balpha$ and $\bbeta$ values close
to zero.
\ourmod\ expresses a more dynamic behavior than FiLM, which could explain
its performance gain.
Next, we remove the conditioning information from either $\balpha$ or $\bbeta$
by replacing it with its mean across the training set (Fig.~\ref{fig:alphabeta}, center).
The larger difference when removing conditioning from $\bbeta$ suggests that
\ourmod\ is more effective in integrating tabular information to
shift feature maps, whereas FiLM is overall less effective and depends on scaling and shifting.
This is also supported by our third test time ablation, where we add
Gaussian noise to $\balpha$ or $\bbeta$ (Fig.~\ref{fig:alphabeta}, right).
For \ourmod, the performance loss is larger when distorting $\bbeta$, whereas
it is equal for FiLM.
In addition, \ourmod\ seems in general more robust to inaccurate $\balpha$ or $\bbeta$
than FiLM.

\begin{table}[t]
  \centering%
  \caption{\label{tab:ablation-study}%
  Test set performance for different configurations.
  The proposed configuration (last row) uses \ourmod\ before the
  first convolution with shift, scale, and $\sigma(x) = x$.}
  \begin{scriptsize}%
  \begin{tabular}{lcccc}%
    \toprule
    Configuration & Balanced Accuracy & Concordance Index \\
    \midrule
    Before Last ResBlock      &      0.598 $\pm$ 0.038 %
    &  0.749 $\pm$ 0.052 \\
    Before Identity-Conv &      0.616 $\pm$ 0.018 %
    &  0.745 $\pm$ 0.036 \\
    Before 1st ReLU      &      0.622 $\pm$ 0.024 %
    &  0.713 $\pm$ 0.085 \\
    Before 2nd Conv      &      0.612 $\pm$ 0.034 %
    &  0.759 $\pm$ 0.052 \\
    \hline
    $\bm{\alpha}_i = \mathbf{1}$  &      0.581 $\pm$ 0.053 %
    &  0.743 $\pm$ 0.015 \\
    $\bm{\beta}_i = \mathbf{0}$   &      0.609 $\pm$ 0.024 %
    &  0.746 $\pm$ 0.057 \\
    \hline
    $\sigma(x) = \mathrm{sigmoid}(x)$              &      0.600 $\pm$ 0.025 %
    &  0.756 $\pm$ 0.064 \\
    $\sigma(x) = \mathrm{tanh}(x)$                 &      0.600 $\pm$ 0.025 %
    &  0.770 $\pm$ 0.047 \\
    \hline
    Proposed        &      0.622 $\pm$ 0.044 %
    &  0.748 $\pm$ 0.045 \\
    \bottomrule
  \end{tabular}%
  \end{scriptsize}
\end{table}

\section{Conclusion}

Brain MRI can only capture a facet of the underlying dementia-causing changes and
other sources of information such as patient demographics, laboratory measurements,
and genetics are required to see the MRI in the right context.
Previous methods often focus on extracting
image information via deep neural networks, but na{\"i}vely account for
other sources in the form of tabular data via concatenation, which
results in minimal exchange of information between
image- and tabular-related parts of the network.
We proposed the Dynamic Affine Feature Map Transform (\ourmod)
to incite or repress high-level concepts learned from
a 3D image, conditional on both image and tabular information.
Our experiments on Alzheimer's disease diagnosis and time-to-dementia prediction
showed that \ourmod\ outperforms previous deep learning approaches
that combine image and tabular data.
Overall, our results support the case that \ourmod\ is a versatile approach
to integrating image and tabular data that is likely applicable to many medical
data analysis tasks outside of dementia too.

\subsubsection*{Acknowledgements.}
This research was supported by the Bavarian State Ministry of Science and the Arts and coordinated by the Bavarian Research Institute for Digital Transformation,
and the Federal Ministry of Education and Research in the call for Computational Life Sciences (DeepMentia, 031L0200A).
\bibliographystyle{splncs04}
\bibliography{paper680}

\end{document}